%% file: ms.tex
\documentclass[aps,prl,twocolumn,superscriptaddress]{revtex4-1}
\usepackage{graphicx}
\usepackage{hyperref} 
\usepackage{amsmath}  
\usepackage{color}
\usepackage[absolute,overlay]{textpos}  
\newcommand{\OH}{OH${}^-$}

\newcommand{\tnu}{\tilde{\nu}}
\newcommand{\icm}{cm$^{-1}$}

\begin{document}
\hyphenation{thermo-metry}


\title{Radiative rotational lifetimes and state-resolved relative detachment
  cross sections from photodetachment thermometry of molecular anions in a
  cryogenic storage ring}


\author{C.~Meyer}\email{Email: cmeyer@mpi-hd.mpg.de}
\affiliation{Max-Planck-Institut f\"ur Kernphysik, D-69117 Heidelberg, Germany}
\author{A.~Becker}
\affiliation{Max-Planck-Institut f\"ur Kernphysik, D-69117 Heidelberg, Germany}
\author{K.~Blaum}
\affiliation{Max-Planck-Institut f\"ur Kernphysik, D-69117 Heidelberg, Germany}
\author{C.~Breitenfeldt}
\affiliation{Max-Planck-Institut f\"ur Kernphysik, D-69117 Heidelberg, Germany}
\affiliation{Institut f\"ur Physik, 
  Ernst-Moritz-Arndt-Universit\"at Greifswald, D-17487 Greifswald, Germany}
\author{S.~George}
\affiliation{Max-Planck-Institut f\"ur Kernphysik, D-69117 Heidelberg, Germany}
\author{J.~G\"ock}
\affiliation{Max-Planck-Institut f\"ur Kernphysik, D-69117 Heidelberg, Germany}
\author{M.~Grieser}
\affiliation{Max-Planck-Institut f\"ur Kernphysik, D-69117 Heidelberg, Germany}
\author{F.~Grussie}
\affiliation{Max-Planck-Institut f\"ur Kernphysik, D-69117 Heidelberg, Germany}
\author{E.~A.~Guerin}
\affiliation{Max-Planck-Institut f\"ur Kernphysik, D-69117 Heidelberg, Germany}
\author{R.~von~Hahn}
\affiliation{Max-Planck-Institut f\"ur Kernphysik, D-69117 Heidelberg, Germany}
\author{P.~Herwig}
\affiliation{Max-Planck-Institut f\"ur Kernphysik, D-69117 Heidelberg, Germany}
\author{C.~Krantz}
\affiliation{Max-Planck-Institut f\"ur Kernphysik, D-69117 Heidelberg, Germany}
\author{H.~Kreckel}
\affiliation{Max-Planck-Institut f\"ur Kernphysik, D-69117 Heidelberg, Germany}
\author{J.~Lion}
\affiliation{Max-Planck-Institut f\"ur Kernphysik, D-69117 Heidelberg, Germany}
\author{S.~Lohmann}
\affiliation{Max-Planck-Institut f\"ur Kernphysik, D-69117 Heidelberg, Germany}
\author{P.~M.~Mishra}
\affiliation{Max-Planck-Institut f\"ur Kernphysik, D-69117 Heidelberg, Germany}
\author{O.~Novotn\'{y}}
\affiliation{Max-Planck-Institut f\"ur Kernphysik, D-69117 Heidelberg, Germany}
\author{A.~P.~O'Connor}
\affiliation{Max-Planck-Institut f\"ur Kernphysik, D-69117 Heidelberg, Germany}
\author{R.~Repnow}
\affiliation{Max-Planck-Institut f\"ur Kernphysik, D-69117 Heidelberg, Germany}
\author{S.~Saurabh}
\affiliation{Max-Planck-Institut f\"ur Kernphysik, D-69117 Heidelberg, Germany}
\author{D.~Schwalm}
\altaffiliation{Deceased.}
\affiliation{Max-Planck-Institut f\"ur Kernphysik, D-69117 Heidelberg, Germany}
\affiliation{Weizmann Institute of Science, Rehovot 76100, Israel}
\author{L.~Schweikhard}
\affiliation{Institut f\"ur Physik, 
  Ernst-Moritz-Arndt-Universit\"at Greifswald, D-17487 Greifswald, Germany}
\author{K.~Spruck}
\affiliation{Max-Planck-Institut f\"ur Kernphysik, D-69117 Heidelberg, Germany}
\affiliation{Institut f\"ur Atom- und Molek\"ulphysik, 
  Justus-Liebig-Universit\"at Gie\ss en, D-35392 Gie\ss en, Germany}
\author{S.~Sunil~Kumar}
\affiliation{Max-Planck-Institut f\"ur Kernphysik, D-69117 Heidelberg, Germany}
\author{S.~Vogel}
\affiliation{Max-Planck-Institut f\"ur Kernphysik, D-69117 Heidelberg, Germany}
\author{A.~Wolf}
\affiliation{Max-Planck-Institut f\"ur Kernphysik, D-69117 Heidelberg, Germany}


\date{\today}

\begin{abstract}
  Photodetachment thermometry on a beam of \OH\ in a cryogenic storage ring
  cooled to below 10~K is carried out using two-dimensional, frequency and time
  dependent photodetachment spectroscopy over 20 minutes of ion storage. In
  equilibrium with the low-level blackbody field, we find an effective radiative
  temperature near 15~K with about 90\% of all ions in the rotational ground
  state.  We measure the $J=1$ natural lifetime (about 193~s) and determine the
  \OH\ rotational transition dipole moment with 1.5\% uncertainty.  We also
  measure rotationally dependent relative near-threshold photodetachment cross
  sections for photodetachment thermometry.
\end{abstract}

\pacs{PACS}

\maketitle



Small molecular ions and their interactions in the lowest rotational states are
crucial for the formation of molecules in interstellar space and for
low-temperature plasma chemistry in general \cite{herbst_can_1981,
  petrie_interstellar_1997}.  Both cations and anions
\cite{mccarthy_laboratory_2006} were observed in space based on rotational
spectroscopy.  While line positions are well studied in the laboratory for many
relevant ions, experimental information on their line intensities is scarce.
Instead, line strengths for ionic rotational transitions are generally obtained
from calculated molecular dipole moments.  This is mainly due to difficulties of
performing absolute laboratory measurements on the dipole moment in the presence
of an ionic charge, small radiative absorption in dilute ionic targets, and long
natural lifetimes of rotational levels.  Recently, cryogenic storage rings for
fast ion beams were taken in operation
\cite{backstrom_storing_2015,von_hahn_cryogenic_2016} that allow the low-lying
rotational levels in small molecular ions to relax by spontaneous emission
toward equilibrium with a low-temperature blackbody radiation field
\cite{oconnor_photodissociation_2016}.  In this Letter, we show that
near-threshold photodetachment can be used to follow the in-vacuo rotational
relaxation over times long compared to the natural lifetime of the first excited
rotational level, and obtain Einstein coefficients for the lowest rotational
transitions of the \OH\ molecular anion.  The measured electric dipole moment
differs significantly from the theoretical values available.

Photodetachment experiments on molecular anions reveal a wealth of information
on their structure and reactive dynamics.  At photon energies close to the
electron binding energy, the detachment cross section is a powerful probe for
the internal states of the anion and the neutral daughter molecule as well as
for the interaction of the outgoing low-energy electron with the neutral
molecule.  With its simple rotational structure in the $^1\Sigma^+$ ground
state, \OH\ has been intensely studied, a particular focus being the strong
deviations from the Wigner threshold law in the photon energy dependence
\cite{engelking_effects_1984,smith_high-resolution_1997}.  Although until now
significant uncertainties remain in predicting the cross sections, rotational
level distributions in the anion have been characterized using the threshold
structure \cite{schulz_oh_1982,goldfarb_photodetachment_2005}.  In a cold
ion-trap environment, near-threshold photodetachment was applied as a method for
rotational thermometry on \OH\ anions under buffer-gas cooling
\cite{otto_internal_2013,hauser_rotational_2015}.  In these studies, the
relative photodetachment cross-sections for the various initial and final state
dependent thresholds were the main uncertainty in deconvoluting the rotational
population fractions of \OH\ from the measured photodetachment rates.  In the
present work we find that probing the radiative rotational relaxation of \OH\
offers a way to obtain the convolution parameters of near-threshold
photodetachment thermometry consistently on an experimental basis.  Hence, in
addition to the rotational lifetime measurements, we also determine relative
photodetachment cross sections over a sample of near-threshold energies for
individual rotational levels of \OH.

In the present experiment, a beam of \OH\ anions from a Cs sputter ion source
(expected rotational temperature of a few thousand K
\cite{menk_vibrational_2014}) is accelerated to 60~keV and injected into the
cryogenic storage ring CSR \cite{von_hahn_cryogenic_2016}.  About $10^7$ ions
are stored at an ambient temperature near 6~K and a residual gas density
corresponding to $<10^{-14}$~mbar room-temperature pressure.  The ion storage
time up to 1\,200~s covers the spontaneous decay of low-lying excited $J$ levels
of \OH\ ($\sim21$\,s and $\sim190$\,s for $J=2$ and $1$, respectively, while
vibrations radiatively decay within $<$1~s \cite{amitay_rotational_1994}).  In a
field-free straight section, laser beams are overlapped with the ion beam in
co-propagating direction at a grazing angle of $3.4^\circ$.  With laser and ion
beam diameters of $\sim7$~mm and $\sim30$~mm, respectively, the interaction zone
is $\sim50$~cm long.  The fast particles neutralized by photodetachment leave
the closed orbit of the CSR and are counted by a large microchannel plate
detector $\sim3$~m downstream from the interaction region.  A continuous-wave
helium--neon (HeNe) laser at 633\,nm with an effective power of 0.7\,mW in the
interaction region yields a steady small rate of photodetachment events.
Doppler-shifted to the \OH\ rest-frame its wavenumber is $\tnu_r=15\,754$~\icm,
which is $\sim$1000~\icm\ above the first \OH($J=0$) detachment threshold
($\tnu_{\rm EA}=14\,741.0$~\icm) corresponding to the ground-state electron
affinity \cite{schulz_oh_1982,goldfarb_photodetachment_2005}.  Measurements with
buffer-gas cooled \OH\ ions \cite{hlavenka_absolute_2009} near 15\,106 \icm\ and
15\,803 \icm\ showed that the photodetachment cross section is independent
(within $\sim$10\%) of the rotational temperature between 8~K and 300~K.  Given
these small variations of the photodetachment cross section with $J$, we use the
photodetachment rate at $\tnu_r$ as a reference signal for the number of \OH\
ions in the laser interaction zone.  The decay of this signal as a function of
time is close to exponential with a time constant of 607(2)~s at $>$150~s (only
$J=0$ and 1 surviving).  By searching for a component in this signal due to the
$J=1$ decay, we find that the relative difference between the photodetachment
cross sections of these two states is $<$3\%.

With the signal at $\tnu_r$ for normalization, we measure the neutral rates from
a second, probing laser.  At similar interaction geometry as the HeNe reference
laser, pulses with $\sim0.5$~mJ, 3--5~ns duration and a repetition rate of
20~s$^{-1}$ are applied by a tunable pulsed optical parametric oscillator (OPO)
laser (EKSPLA NT342B).  Close to the excited-$J$ photodetachment thresholds, up
to 10 probing wave numbers $\tnu_k$ ($k=1\dots10$, given in the \OH\ rest-frame)
are used.  The neutrals reach the detector within $\sim$5~$\mu$s after the laser
pulses, reflecting the particle flight times.  Their counts in a suitable delay
window are accumulated as the signal $N(\tnu_k,t)$, where $t$ is the time after
the ion beam injection.  Similarly, the counts due to the HeNe laser are
accumulated during the breaks between the laser pulses and recorded as the
reference $N(\tnu_r,t)$.  Laser pulsing starts few ms after injection. Probing
wavenumbers $\tnu_k$ are cycled through with typically 100 laser pulses at a
single value. The ion beam was dumped 31~s, 300~s or 1200~s after injection.
For a run, many of such injections and observation periods were repeated.  The
starting value of the $\tnu_k$ cycle was varied to realize a two-dimensional
spectroscopy scheme that applies all probing wave numbers to all beam storage
times with a suitable time binning.

\begin{figure}
\includegraphics[width=8.4cm]{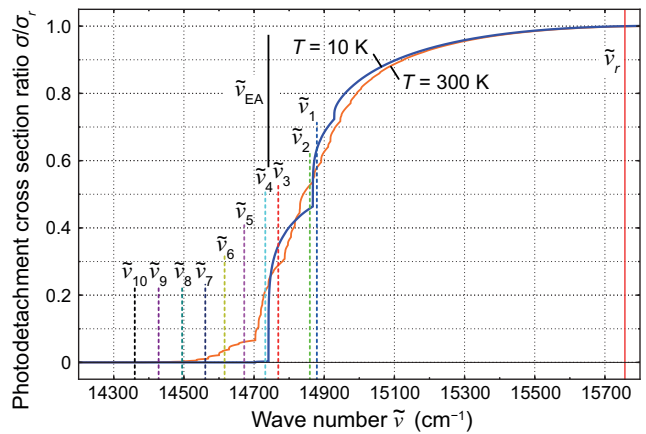}
\caption{Near-threshold photodetachment cross section of \OH\ modeled by
  $\sigma_T(\tnu;T)=\sum_Jp_J(T)\sigma_J(\tnu)$ with $\sigma_J(\tnu)$ for
  $a=0.20, b=-2.8$ and for $T=10~K$ and 300~K as labeled.  The reference cross
  section is $\sigma_r=\sigma_{J=0}(\tnu_r)$ at the HeNe laser wave number (full
  vertical mark; see the text).  Probing wave numbers $\tnu_k$ are indicated by
  dashed marks.\label{fig:probing}}
\end{figure}

\begin{figure*}[t]
\includegraphics[width=17.5cm]{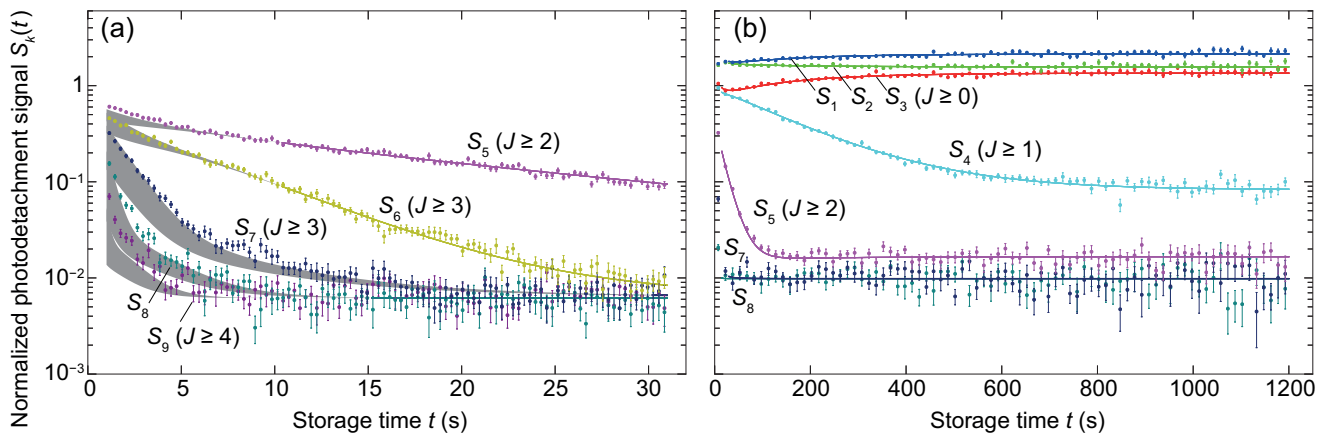}
\caption{Photodetachment signals $S_k(t)=N(\tnu_k,t)/N(\tnu_r,t)$ as functions
  of the ion storage time for two sample runs with 1151 (a) and 38 (b)
  injections.  Ranges of $J$-levels contributing to the signals are marked.
  Data points represent the measured count ratios and their $1\sigma$
  uncertainties.  The full lines show a fit to the signals of all runs using the
  modeled photodetachment cross sections and level populations.  In (a) the
  shaded areas indicate the model variation with $T_0$ at the short times when
  the data were excluded from the fit (see text).\label{fig:data}}
\end{figure*}

For the design of the experiment and the basic understanding of its results,
modeling of \OH\ near-threshold photodetachment \cite{hotop74,schulz_oh_1982}
was crucial.  Starting from \OH($J$), $s$-wave photodetachment can leave the OH
radical ($X\,^2\Pi$) in up to eight final levels, each forming a threshold $j$
at wave number $\tnu_j$. (For $J=0$ only three thresholds are allowed.)  The
near-threshold detachment rate is described \cite{schulz_rotational_1983} by a
Wigner-type dependence $I_j(\tnu-\tnu_j)^a$ where, in laser measurements just
above threshold, appropriate values of $a$ were found to vary from one threshold
to the next with a typical range of $a\approx0.20\ldots0.25$.  The threshold
intensities $I_j$ follow from the angular momenta and the OH fine structure
mixing \cite{schulz_rotational_1983, goldfarb_photodetachment_2005}, in
reasonable agreement with the observations.  Much less is known about the
validity of the threshold power law at higher above-threshold photon energies
($\tnu-\tnu_j$), which in photodetachment thermometry can reach up a few hundred
\icm.  Our data analysis is largely independent of a photodetachment cross
section model and we obtain the $J$-dependent relative cross sections from the
time dependence at the various probing wave numbers $\tnu_k$. We have to relate
only one of the $J$-dependent cross sections at a single $\tnu_k$ to the
reference rate via the cross section model. Moreover, unresolved contributions
of higher $J$ ($\geq4$) are included based on the model results. Specifically
the analytical model for a threshold $j$ is chosen as
$\sigma_j(\tnu)=\hat{I}_j(\tnu-\tnu_j)^a(\tnu/\tnu_j)^{b-a}$ with
$\hat{I}_j=I_j/(2J+1)$ \cite{supplement}.  
\nocite{rudmin_structure_1996}\nocite{maillard_highresolution_1976}
\nocite{Gewurtz1975}\nocite{breyer_high_1981}\nocite{bernath_spectra_2005}
\nocite{chang_cmistark_2014}\nocite{werner_molecular_1983}
\nocite{molnar_appendixes_2011}
A factor of power $b-a$ is introduced
to model cross section deviations from the threshold law at higher $\tnu$.  Only
few data \cite{branscomb_photodetachment_1966,hlavenka_absolute_2009} are
available for the cross section in this $\tnu$ range and indicate a maximum at,
roughly, 16\,000 \icm.  We choose $a=0.20$ and $b=-2.8$ for a model
approximating the photon energy dependence.

The cross section model for \OH($J$) is obtained as
$\sigma_{J}(\tnu)=\sum_{j(J)}\sigma_j(\tnu)$ where $j(J)$ denotes all thresholds
for a given $J$.  For \OH\ with a rotational temperature $T$ the cross section
(Fig.\ \ref{fig:probing}) is the average over $\sigma_{J}(\tnu)$ with the
population probabilities $p_J$.  For $T=10$~K, it is dominated by $J=0$
($p_0=0.987$) with a rise at $\tnu_{\rm EA}$ followed by two further thresholds
within the next 200~\icm.  Higher $J$ levels populated at $T=300$~K lead to a
broadening of the threshold structures, while from $\sim600$~\icm\ above
$\tnu_{\rm EA}$ the cross sections are largely independent of $T$ and of $J$.

Among the probe wave numbers, $\tnu_1\ldots\tnu_3$ lie well above
$\tnu_{\rm EA}$ (Fig.\ \ref{fig:probing}) and, thus, yield contributions from
all $J\geq0$.  Other $\tnu_k$ successively exclude low-$J$ levels; contributing
$J$'s are, e.g., $J\geq1$ for $\tnu_4$ and $J\geq2$ for $\tnu_5$.  Signals
$S_k(t)=N(\tnu_k,t)/N(\tnu_r,t)$, shown in Fig.\ \ref{fig:data}, are obtained by
normalizing the counts $N(\tnu_k,t)$ to the reference $N(\tnu_r,t)$.  A short
run up to 31~s, using $\tnu_5\ldots\tnu_9$, shows the successive cooling of
higher rotational levels.  After the fast decay of $J\geq4$ ($t\gtrsim10$~s),
$S_6$ essentially shows the relaxation of $J=3$.  Similarly, $S_5$ at later
times ($\gtrsim30$~s) represents the $J=2$ decay.  In a long run up to 1\,200~s,
$S_4$ for $t\gtrsim200$~s represents the pure $J=1$ decay.  Furthermore, the
signals $S_1 \ldots S_3$ clearly show the different $J$ contributions by their
characteristic time dependences consistent with $S_4$ and $S_5$, in particular.

The normalized signals represent a convolution of the time-dependent $J$-level
populations $p_J$ with a matrix representing the relative photodetachment cross
sections at the probing wave numbers:
$S_k(t)=S_0\sum_J p_J(t)\phi_k\sigma_J(\tnu_k)/\sigma_r$.  Factors $\phi_k$
close to 1 describe the small relative variations of the OPO laser flux over
$\tnu_k$ (known within $\sim$3\%).  The OH$^-$ radiative relaxation can be well
described by only a few parameters: the Einstein coefficients $A_J$ for
transitions $J \rightarrow J-1$, the photon occupation numbers $n(\tnu_J)$ of
the ambient radiation field at the transition energies $\tnu_J$ for transitions
$J \rightarrow J+1$, and the initial populations $p_J^0$ of the rate model
\cite{supplement}.  An overall scaling parameter $S_0$ (in the range of $3.4$ to
$4.4$) takes into account the integrated powers and the slightly different
overlaps of the two laser beams.

The signals $S_k(t)$ were fitted by a single model simultaneously for all runs.
We assume that the $n(\tnu_J)$ are given by a radiative temperature $T_r$
according to Bose-Einstein statistics.  The populations $p_J^0$ from the
excitation in the ion source are described by a temperature $T_0$.  We
independently varied in the fit all $A_J$ for $J\leq3$.  At the various
$\tnu_k$, the signal variations differently reflect the radiative time constants
and the $J$-dependent photodetachment cross sections
$\sigma_J(\tnu_k)/\sigma_r$.  Hence, the relative cross sections can also be
obtained.

We fit $S_{1,2}$ for $t>30$~s and $S_3$ to $S_6$ for $t>10$~s to determine
relative $\sigma_J$ values for $J=0$ to 3.  The time limits are set to safely
ensure the decay of higher-$J$ levels. On the other hand, the radiative
lifetimes of $J=1$ to 3 are sensitively probed.  For fitting $S_0$, one of the
relative photodetachment cross sections, for which we chose
$\sigma_{J=0}(\tnu_3)/\sigma_r$, has to be set to its calculated value.  In
$S_7$ to $S_{10}$, contributions from various higher $J$ overlap at shorter
times.  We do not fit these short-time data, but only compare them to the model
using the calculated $\sigma_J(\tnu_k)/\sigma_r$.  The higher-$J$ lifetimes are
set according to the relation
$A_J=16\pi^3\tnu_{J-1}^3\mu_0^2J/3\epsilon_0h(2J+1)$ \cite{bernath_spectra_2005}
using the dipole moment $\mu_{0,J\geq3}$ from the fitted $J=3$ lifetime.  The
backgrounds in these data are fitted for $t>15$~s ($t>30$~s for $S_7$).
Separate fits were performed setting the start temperature $T_0$ between 1000~K
and 6000~K. Within the fitted time ranges the model curves and the fitted
parameters remain essentially unchanged.  In the short-time ranges excluded from
the fits the results vary significantly.  This is indicated by shaded areas in
Fig.\ \ref{fig:data}(a).  Their upper edges, representing the model for
$T_0=6000$~K, yield best agreement with the data.  Hence, we give the results
for the fit at $T_0=6000$~K and consider the estimated parameter variations over
a range of $\pm$2000~K in $T_0$ as a systematic uncertainty.  The reduced
mean-squared residuals $\chi^2_r$ were close to 1.30 in all cases.

\begin{figure}[t]
\includegraphics[width=8.4cm]{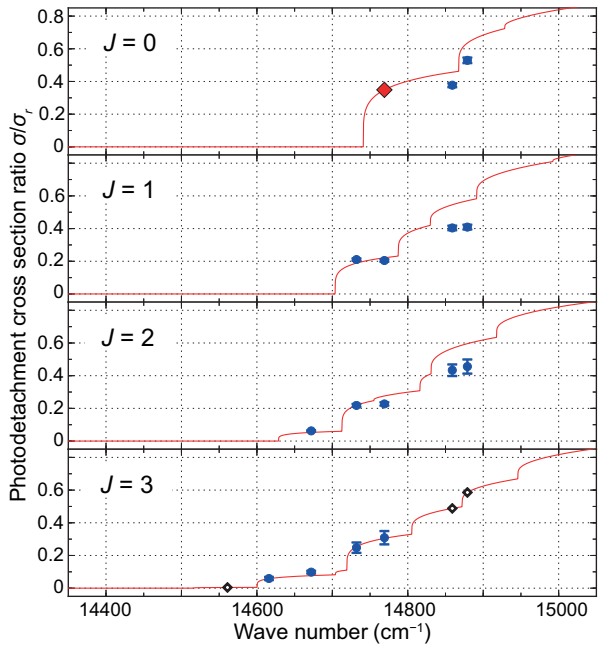}
\caption{Near-threshold photodetachment cross section ratios
  $\sigma_J(\tnu)/\sigma_r$ of \OH\ for the probing lines ($\tnu=\tnu_k$)
  determined from the fit (filled symbols with overall $1\sigma$, statistical
  and systematic uncertainties) and the model with $a=0.20, b=-2.8$ (full
  lines).  Open diamonds mark theoretical values of $\sigma_J(\tnu_k)/\sigma_r$
  for the channels where separate features cannot be identified in the data.
  The model value of $\sigma_{J=0}(\tnu_3)/\sigma_r$ serving as a reference (see
  text) is marked by the large diamond.  \label{fig:rel_cs}}
\end{figure}

Two further experimental features were included in the fit.  First, a small
fraction of $^{17}$O$^-$ ions of $(1.7\ldots5.9) \times10^{-3}$
\cite{supplement} occured in the stored \OH\ beam and, for the O$^-$ electron
affinity near 11\,784.7 \icm\ \cite{blondel_electron_2005}, led to a small
constant neutralization background even for the probing lines far below the \OH\
threshold.  It was considered as a constant background in $S_k(t)$ with
independent values for different runs. Moreover, reflection of laser light at
the downstream vacuum window produced a small additional probing-light component
Doppler-shifted by $\tnu'_k-\tnu_k \approx +80$~\icm.  This yields a small
contribution from $J=0$ ions in $S_5$ (probing $J\geq2$) slightly shifting the
$S_5$ equilibrium level in Fig.\ \ref{fig:data}(b).  The weak, Doppler-shifted
reflected lines ($\tnu'_k$) are consistently included in the model (with
theoretical $\sigma_J(\tnu'_k)/\sigma_r$) and the reflection factors (up to a
few percent) are allowed to vary between different runs.  In the fit results,
systematic uncertainties due to the starting temperature $T_0$ have a
few-percent effect on the relative cross sections, while they are neglible for
the radiative decay and the equilibrium population.  The estimated limits for
rotational variations of the reference photodetachment cross section and
differences in depleting the various $J$-levels by laser photodetachment
introduce \cite{supplement} systematic uncertainties on the radiative decay
rates which are included in the overall uncertainty limits of the results.

The fits shown in Fig.\ \ref{fig:data} yield precise details on the radiative
equilibrium in the low-level blackbody field of the CSR, the relative probing
cross sections and the radiative decay rates of the OH$^-$ ion.  From the fitted
photon occupation number at the $J=0\to1$ transition frequency in \OH\
(37.47~\icm), we obtain an effective radiative temperature of
$T_r=15.1(1)$~K. Considering the CSR vacuum-chamber temperature
\cite{von_hahn_cryogenic_2016} near 6~K, the effective relative contribution
from room-temperature surfaces to the radiation field is found \cite{supplement}
to be $5.7(2)\times10^{-3}$.

The fit results for the state-selected relative photodetachment cross sections
at the various probe wave numbers are shown in Fig.\ \ref{fig:rel_cs}.  It is
clearly visible that the simple model is inappropriate for the probing energies
of 100~\icm\ or more above the lowest thresholds in $J=0$ to 2.  As a simple
change of the common exponent $a$ of all threshold laws does not improve the
overall agreement, the results call for more detailed cross-section calculations
beyond simple power-law models.  But the experimental cross section ratios (see
listing in Ref.\ \cite{supplement}) can directly serve for future
photodetachment thermometry, deducing rotational populations from relative
photodetachment rates on a theory-independent basis. This will include ion trap
experiments such as those of
Refs.\ \cite{otto_internal_2013,hauser_rotational_2015}, where photodetachment
thermometry serves to study cold inelastic collisions by laser depletion of
rotational levels.

\begin{table}[t]
  \caption{Einstein coefficients, natural lifetimes $\tau_J$ and the
    corresponding transition dipole moments measured for \OH\ rotational 
    states with the combined statistical and systematic uncertainty
    ($1\sigma$ confidence level).\label{tab:fit_values}}
\begin{ruledtabular}
\begin{tabular}{ccccl}
  & $J=1$ & $J=2$ & $J=3$ &  \multicolumn{1}{c}{unit} \\ 
  \hline
  $A_J$ & 0.00517(18) & 0.0478(48) & 0.189(13) &  s$^{-1}$\\
  $\tau_J=A_J^{-1}$ & 193(7) &   20.9(2.1) & 5.30(37)  &  s \\
  $\mu_0$ & 0.970(17) & 0.952(48) &  0.997(35) 
                                    \footnote{Value used for $\mu_{0,J\geq3}$} & D \\
\end{tabular}
\end{ruledtabular}
\end{table}

Results for the Einstein coefficients and the molecular dipole moment are listed
in Table \ref{tab:fit_values}. To our best knowledge, direct in-vacuo lifetime
measurements on low-lying, purely rotationally excited states in small molecules
have not been reported previously.  As expected at the given accuracy, the
molecular dipole moments extracted from $A_J$ assuming the elementary
H\"onl-London factors are compatible among each other within experimental
uncertainties.  The weighted average of $\mu_0=0.982(15)$~D can be compared to
calculations of the \OH\ dipole moment.  Values at the equilibrium internuclear
distance are $1.050$ to $1.072$~D in earlier \cite{werner_molecular_1983} and
1.10~D in recent work \cite{vamhindi_accurate_2016}. Taking the dipole moment
function of Ref.\ \cite{werner_molecular_1983}, vibrational averaging reduces
$\mu_0$ by at most 0.04~D ($\mu_0=1.037$~D \cite{supplement}).  Hence, we find
that current theory overestimates the \OH\ dipole moment by $(5.3 \pm 1.5)\%$
and underestimates the \OH\ rotational lifetimes by about $(10\pm3)\%$.

In the described time-dependent near-threshold photodetachment spectroscopy
using a cryogenic storage ring, the well-understood dynamics of in-vacuo
radiative relaxation allowed us to clearly identify the contributions of single
rotational levels.  This single-level sensitivity will be useful in the future
to probe rotational population changes by in-ring molecular collisions.
Moreover, a method was demonstrated to perform precise laboratory measurements
of natural lifetimes and line intensities for extremely slow, purely rotational
transitions in molecular ions.  Rotational lifetimes from such measurements add
a further, so far unavailable experimental benchmark for quantum chemical
calculations.  Further improvements of the accuracy and applications to anionic
molecules with more complicated rotational level structures can be envisaged.

Support by the Max Planck F\"orderstiftung is gratefully acknowledged.  F.~G.,
E.~A.~G., A.~P.~O. and H.~K. were supported by the European Research Council
under Grant Agreement No.\ StG~307163.



%

\end{document}


%
{\raggedright%
  \begin{minipage}[b]{\textwidth} \bfseries{\small{Online Supplementary Material}\par}\vspace{5mm} \bfseries{
  {Radiative rotational lifetimes and state-resolved relative detachment cross sections from photodetachment thermometry of
  molecular anions in a cryogenic storage ring}%
    }\par
  \end{minipage}\\[8mm]
  \begin{minipage}[b]{\textwidth} \raggedright \small 
    C.~Meyer$^1$, A.~Becker$^1$, K.~Blaum$^1$, C.~Breitenfeldt$^{1,2}$, S.~George$^1$, J.~G\"ock$^1$, M.~Grieser$^1$, F.~Grussie$^1$,
    E.~A.~Guerin$^1$, R.~von~Hahn$^1$, P.~Herwig$^1$, C.~Krantz$^1$, H.~Kreckel$^1$, J.~Lion$^1$, S.~Lohmann$^1$, P.~M.~Mishra$^1$,
    O.~Novotn\'{y}$^1$, A.~P.~O'Connor$^1$, R.~Repnow$^1$, S.~Saurabh$^1$, D.~Schwalm$^{1,3}$, L.~Schweikhard$^2$, K.~Spruck$^{1,4}$,
    S.~Sunil~Kumar$^1$, S.~Vogel$^1$, and A.~Wolf$^1$
  \end{minipage}\\[3mm]%
  \begin{minipage}{\textwidth} \raggedright\itshape \footnotesize%
    $^1$Max-Planck-Institut f\"ur Kernphysik, 69117 Heidelberg, Germany; %
    $^2$Institut f\"ur Physik, Ernst-Moritz-Arndt-Universit\"at Greifswald, D-17487 Greifswald, Germany; %
    $^3$Weizmann Institute of Science, Rehovot 76100, Israel; %
    $^4$Institut f\"ur Atom- und Molek\"ulphysik, Justus-Liebig-Universit\"at Gie\ss en, D-35392 Gie\ss en, Germany
  \end{minipage}
  \par}
\vspace{2mm}
\hrule
\vspace{10mm}
\thispagestyle{empty}
\def\theequation{S\arabic{equation}}
\def\thefigure{S\arabic{figure}}
\def\thetable{S\Roman{table}}
\bibliographystyle{apsrev4-1}

\newcommand{\Ox}{O${}^-$}   
\newcommand{\OH}{OH${}^-$}
\newcommand{\CH}{CH${}^+$}
\newcommand{\tnu}{\tilde{\nu}}
\newcommand{\Enu}{\tilde{E}}
\newcommand{\icm}{cm$^{-1}$}

In this Supplementary Material, we collect additional information about the methods of the measurement and data analysis.  In the final
section, we make some results available in numerical form.

\vspace{10mm}
\section*{Photodetachment modeling}

Our modeling of the \OH\ near-threshold photodetachment cross sections is based on the work by \citet{schulz_rotational_1983},
\citet{rudmin_structure_1996}, and \citet{goldfarb_photodetachment_2005}.  Earlier research on the topic is cited in these
publications.  The \OH\ anion is assumed to be in the vibrational ground state.  Its state is specified by the rotational quantum
number $J''$ (denoted by $J$ in the main article) and the parity $(-1)^{J''}$.  In the neutral OH product, two levels exist for a given
total angular momentum $J'$ and parity.  These levels arise from the fine-structure interaction and the angular momentum coupling in
the $X\,^2\Pi$ state of OH.  The energetically lower levels for a given $J'$ are denoted as the F1 levels and the higher ones as the F2
levels (a number parameter $\alpha=1,2$ for F1 and F2, respectively, is used as well).  The intensity of the threshold (wave number
$\tnu_i$) is described \cite{schulz_rotational_1983,goldfarb_photodetachment_2005} by an expression $I_j$ containing the squares of a
linear combination of relevant transition matrix elements, where $j$ represents the set of labels $(J',\alpha,J'')$ for a specific
threshold.  As discussed in the main text, $I_j$ is multiplied by a threshold law of form $(\tnu-\tnu_j)^a$ with a positive fractional
exponent $a$ to describe the energy dependence of the photodetachment intensity.  We choose to model the photodetachment cross section
of a threshold $j$ by $\sigma_j(\tnu)=\hat{I}_j(\tnu-\tnu_j)^a(\tnu/\tnu_j)^{b-a}$ with the parameters $a=0.20\ldots0.25$ and $b=-2.8$
discussed in the main text.  The cross-section intensity factors $\hat{I}_j$ are related to the threshold intensity $I_j$ calculated in
the literature \cite{schulz_rotational_1983,goldfarb_photodetachment_2005} as
\begin{equation}
  \hat{I}_j=I_j/(2J''+1)   \label{eq:crossec}
\end{equation}
considering the average over degenerate states for the initial level \OH$(J'')$. The threshold wave numbers $\tnu_j$ are considered below. 

\begin{table}[t]
  \caption{Spectroscopic constants applied in the model calculations for OH 
    (Table~IV of Ref.\ \cite{maillard_highresolution_1976}) and OH$^-$ (Table~II of Ref.\
    \cite{schulz_oh_1982}; in \icm) in the ground electronic and vibrational levels.\label{tab:const}}
\begin{minipage}{10cm}
\begin{ruledtabular}
\begin{tabular}{rdd}
  & \multicolumn{1}{c}{OH} & \multicolumn{1}{c}{OH$^-$} \\
  \hline
  $\Enu_{\rm EA}$ & 14\,741.0 \footnote{Rounded value keeping agreement with Refs.\ \cite{schulz_oh_1982,goldfarb_photodetachment_2005}.}&\\
  $A_0$ & -139.18 &  \\
  $B_0$ & 18.550 & 18.741 \\
  $D_0$ & 1.916 \times 10^{-3} & 2.052 \times 10^{-3} \\
  $P(0)$ & 0.234 & \\
  $Q_0(0)$ & -0.039 & \\
\end{tabular}
\end{ruledtabular}
\end{minipage}
\end{table}

In the intermediate-coupling model used, the OH energy levels are described by a superposition of Hund's case (a) states
$^2\Pi_{\Omega'}$ where $\Omega'$ is the absolute value of the projection of the total electronic angular momentum on the internuclear
axis.  Following the well established theory for diatomic hydrides with a $^2\Pi$ ground state, with particular reference to the
treatment of HF$^+$ (isoelectronic to OH) by \citet{Gewurtz1975}, and using the molecular constants \cite{maillard_highresolution_1976}
listed for OH in Table~\ref{tab:const}, the energies of the F1 and F2 states with both parities ($e,f$) are obtained by ($\Lambda=1$ in
Ref.\ \cite{Gewurtz1975})
\begin{eqnarray}
\Enu^{{\rm OH}}_{J',1}=\Enu_{\rm F1} & = & B_0[(J'+1/2)^2-1-X/2] - D_0 J'^4 \pm \Delta\Enu_{fe}^{(1)}/2 , \nonumber \\
\Enu^{{\rm OH}}_{J',2}=\Enu_{\rm F2} & = & B_0[(J'+1/2)^2-1+X/2] - D_0 (J'+1)^4 \pm \Delta\Enu_{fe}^{(2)}/2 ,
\end{eqnarray}
where $Y=A_0/B_0$ ($<0$ for OH), and
\begin{equation}
X = [4(J'+1/2)^2 + Y(Y-4)]^{1/2} .
\end{equation}
The lambda doubling is taken into account by the parameters \cite{Gewurtz1975}
\begin{equation}
\Delta\Enu_{fe}^{(1,2)}=
\mp (J'+1/2)\{(\mp1+2/X-Y/X)[P_0/2+Q_0(0)] + 2Q_0(0)(J'+1/2)(J'+3/2)/X\},
\end{equation}
where the upper (lower) signs hold for the indices 1~(2), respectively, and $\Delta\Enu_{fe}=\Enu_f-\Enu_e$.  The $e$ states have
parity $(-1)^{J'-1/2}$ and the $f$ states parity $-(-1)^{J'-1/2}$.  The lowest OH level is $\Enu^{{\rm OH}}_0=\Enu^{{\rm OH}}_{3/2,1}$
with odd parity (i.e., the $e$ level).  The mixing coefficient
\begin{equation}
\kappa=[(X-2+Y)/2X]^{1/2}   \label{eq:kappa}
\end{equation}
governs the amplitudes of the Hund's-case-(a) (fixed $\Omega'$) states in the F1 and F2 levels, further discussed below.  In the case
of OH with $Y<0$ the lower-energy (F1) states have dominantly character $\Omega'=3/2$ and the higher-energy (F2)
states dominantly $\Omega'=1/2$.

The threshold photon energy $\tnu_j$ is obtained from the \OH\ and OH level energies and the electron affinity $\Enu_{\rm EA}$  (see
Table~\ref{tab:const}) as
\begin{equation}
  \nu_j=\Enu^{{\rm OH}}_{J',\alpha}-\Enu^{{\rm OH}}_0+\Enu_{\rm EA}-\Enu^{{\rm OH}^-}_{J''}. \label{eq:ohrotene}
\end{equation}
The \OH\ energy levels, considering the molecules to be in the vibrational ground state, are described by the linear non-rigid rotor
model using
\begin{equation}
\Enu^{{\rm OH}^-}_{J''} =B_0 J''(J''+1)- D_0 J''^2(J''+1)^2.
\end{equation}
For the dominant $s$ wave photodetachment only considered here, the required OH parity is $-(-1)^{J''}$, and $\Enu^{{\rm
    OH}}_{J',\alpha}$ refers to the level of this parity.  For $J''=0$ the lowest $s$-wave threshold leads to the level $\Enu^{{\rm
    OH}}_{3/2,1}$ with odd parity and hence the wave number of this threshold equals $\Enu_{\rm EA}$.

The calculation of the intensities $I_j$ proceeds by first considering the photoexcitation of an intermediate state (complex) of
negative total charge with the angular momentum quantum number $J_C$.  This state evolves to possible final states of OH with labels
($J',\alpha$) and an $s$-wave electron by recoupling of angular momenta and their projection on the internuclear axis.  The angular
momentum $J_C$ follows the dipole transition selection rules ($J_C=J''$ or $J''\pm1$) while the spin 1/2 carried away by the outgoing
electron then allows $J'=J''\pm1/2$ or $J''\pm3/2$ ($J_C=J'\pm1/2$).  The $J_C$ dependence of the phototransition is expressed by a
constant radial transition dipole moment and the H\"onl-London factors.  In the intensity factor, pathways corresponding to different
$J_C$ and $\Omega'$ interfere as expressed by appropriate amplitudes for each ($J',\alpha$) final state.

\begin{figure}
\includegraphics[width=0.5\columnwidth]{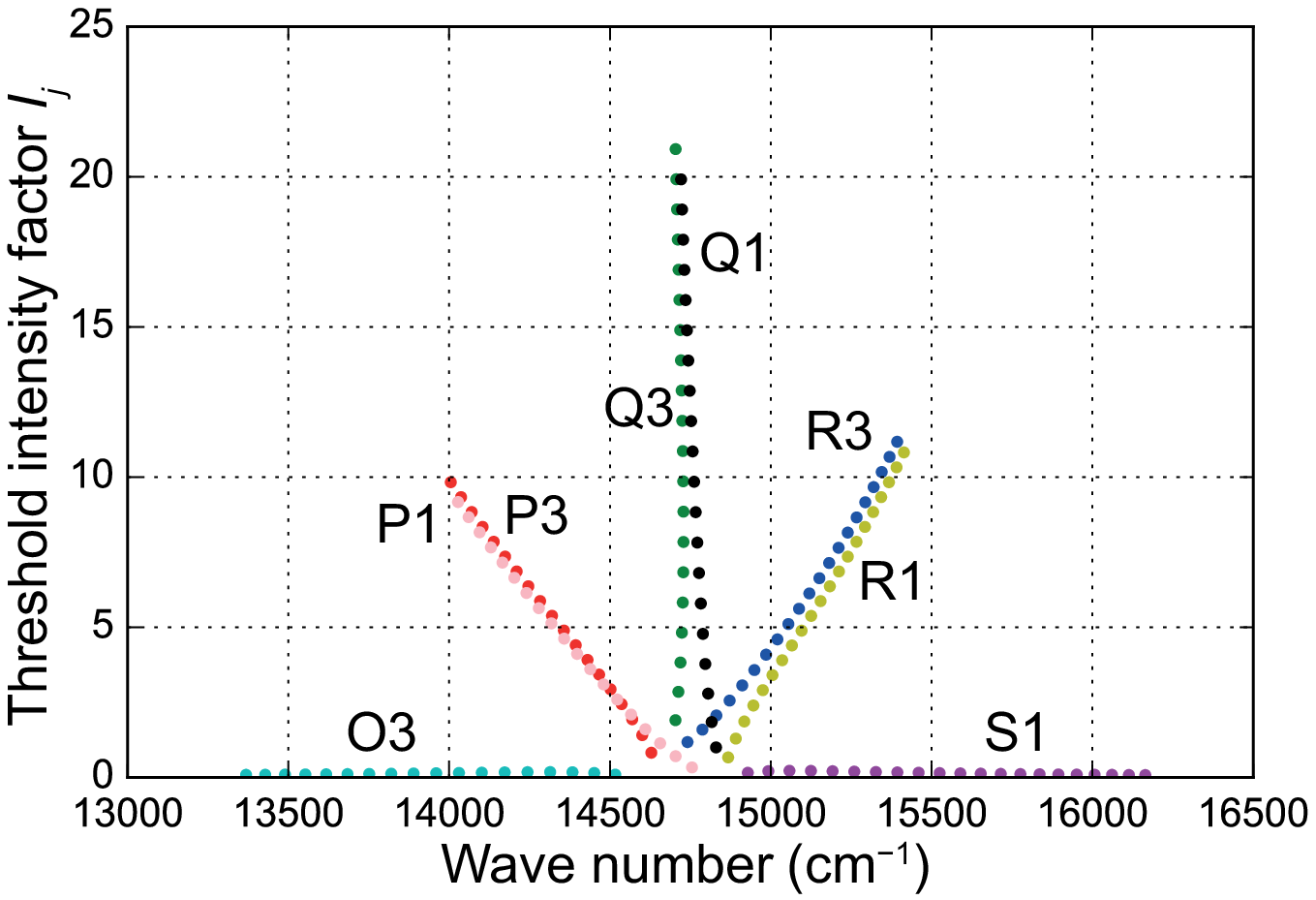}
\caption{Intensity factors $I_j$ for the $s$-wave photodetachment thresholds of the \OH\ rotational levels up to $J''=20$.\label{fig:thres}}
\end{figure}

In their Eqs.~(3) to (6) \citet{goldfarb_photodetachment_2005} correct a small typographic error of Ref.\ \cite{schulz_rotational_1983}
and list the results of the calculation.  We use these expressions setting $I_j=I$ for the quantum numbers corresponding to threshold
$j$.  Here, mixing coefficients $c_1$ and $c_2$ specify the final levels ($J',\alpha$) in terms of their Hund's-case-(a) components.
Specifically, the state amplitudes are represented by $c_1$ for the basis state $^2\Pi_{1/2}$ and by $c_2$ for the basis state
$^2\Pi_{3/2}$.  \citet{goldfarb_photodetachment_2005} emphasize the importance of a proper phase choice for these mixing coefficients
and mention a check regarding the intensities of the various branches $j$.  For the state amplitudes expressions are given by
\citet{schulz_rotational_1983}, Eq.~(5), which regarding absolute values correspond to the coefficient given in our Eq.\
(\ref{eq:kappa}).  However, we found it essential to verify the branch intensities regarding the relative signs of $c_1$ and $c_2$.
For our model calculations, the state amplitudes were then, for an F1 ($\alpha=1$) level, set as
\begin{equation}
  c_1=c_1^{(F1)}=\kappa,~~~~c_2=c_2^{(F1)}=-\sqrt{1-\kappa^2}  \label{eq:mix1} 
\end{equation}
while for an F2 ($\alpha=2$) level,
\begin{equation}
c_1=c_1^{(F2)}=\sqrt{1-\kappa^2},~~~~c_2=c_2^{(F2)}=\kappa   \label{eq:mix2}
\end{equation}
with $\kappa$ from Eq.\ (\ref{eq:kappa}).  Both equations apply to the OH case ($Y<0$).  The branch intensities $I_j$ obtained with
this definition are shown in Fig.\ \ref{fig:thres} and, as discussed below, fulfil the criterium formulated by
\citet{goldfarb_photodetachment_2005}.  Hence, we find that, for OH and for the formula set of Eqs.~(3) to (6) in Ref.\
\cite{goldfarb_photodetachment_2005}, the $c_1, c_2$ coefficients must be chosen with {\it opposite} signs for the F1 state (i.e., the
$\Omega'=3/2$ dominated state).  With relation to \citet{schulz_rotational_1983}, Eq.~(5), we have $c_{1,2}^{(F1)}=c_{1a,2a}$ and
$c_{1,2}^{(F2)}=c_{1b,2b}$ and, unlike quoted in this reference for the OH case, have to use the {\it upper} signs in these
expressions.

In Fig.\ \ref{fig:thres} we show the values of $I_j$ obtained with Eq.~(3) to (5) from \citet{goldfarb_photodetachment_2005} and our
Eqs.\ (\ref{eq:mix1},\ref{eq:mix2}).  As functions of the threshold energy $\tnu_j$, the intensities are clearly grouped in branches.
Their customary notation refers to the quantum number $N'$ (total angular momentum except spin) of the respective dominant component
(i.e., $N'=J'-1/2$ for F1 levels with $\Omega'=3/2$ main component and $N'=J'+1/2$ for F2 levels).  For a Q branch, $N'=J''$, while
branches R and P are defined by $N'=J''\pm1$ and S and O branches by $N'=J''\pm2$, respectively.  Branches denoted as O3, P3, Q3 and R3
lead to F1 levels ($\Omega=3/2$ dominant), while the branches P1, Q1, R1 and S1 lead to F2 levels ($\Omega=1/2$ dominant).  It is seen
that, as expected, the P, Q and R branches of the F1 and F2 levels show similar behavior for high $J''$, while the O3 and S1 branches
keep low amplitudes for all $J''$.  This dramatically changes when the phases in Eqs.\ (\ref{eq:mix1},\ref{eq:mix2}) are reversed.

The values of $I_j$ plotted in Fig.\ \ref{fig:thres} show an approximately linear increase with $J''$.  Hence, at excess photon
energies $\tnu-\tnu_0$ large compared to the spacing between the rotational thresholds $\tnu_j$, we will have a similar near-linear
increase also for $I_j(\tnu-\tnu_j)^a\approx I_j(\tnu-\tnu_0)^a$.  Only the division by the initial-state statistical weight, using the
factors defined in Eq.\ (\ref{eq:crossec}), avoids a drastic dependence on $J''$ of the cross section at high photon energies above the
threshold.  In previous applications \cite{schulz_rotational_1983,goldfarb_photodetachment_2005} the intensities $I_j$ were multiplied
with pure Boltzmann factors (as opposed to level population fractions) in order to obtain the $J''$-dependent relative photodetachment
intensities, which is consistent with our cross-section definition.

\begin{figure}
\includegraphics[width=0.4\columnwidth]{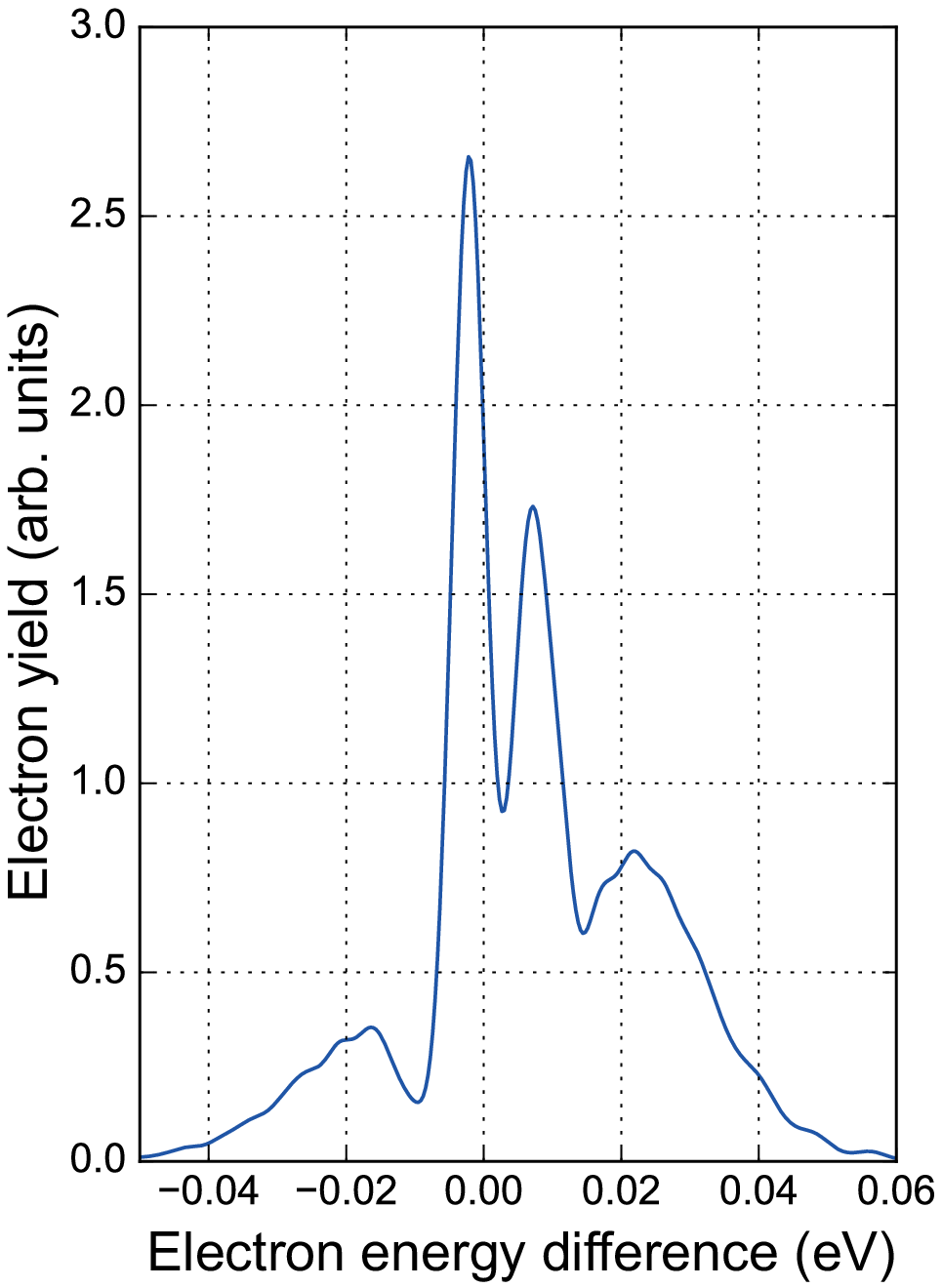}
\caption{Reproduction of the electron spectrum measured by \citet{breyer_high_1981} for \OH\ photodetachment at $\tnu=20\,492$~\icm.
  Comparison to Fig.~1 of Ref.\ \cite{schulz_rotational_1983} shows that the result of \citeauthor{schulz_rotational_1983} is
  precisely reproduced by our implementation of the model.\label{fig:breyer}}
\end{figure}

For another check, we have followed \citet{schulz_rotational_1983} and reproduced their model of the electron spectra measured by
\citet{breyer_high_1981} for high excess photon energies (Fig.\ \ref{fig:breyer}).  It turned out that it is essential to apply the
sign convention of Eqs.\ (\ref{eq:mix1},\ref{eq:mix2}), together with a suitable cross-section definition, to obtain agreement with
both the experimental and the earlier model data.  The model results given in Ref.\ \cite{schulz_rotational_1983} can be reproduced
accurately with the sign convention of Eqs.\ (\ref{eq:mix1},\ref{eq:mix2}), which thus seems to be the one used by these authors (in
spite of the opposite statement near Eq.~(5) of Ref.\ \cite{schulz_rotational_1983}).

\vspace{10mm}
\section*{Radiative relaxation}
Molecules with infrared active modes like \OH\ couple to the surrounding radiative field by emitting and absorbing photons.  The rates
of such processes can be described by the Einstein coefficients.  We assume that the molecules are in the vibrational ground state
only.  Denoting in this section the \OH\ quantum number by $J$, the level energies $\Enu_{J} = \Enu^{{\rm OH}^-}_{J''}$ are given by
Eq.\ (\ref{eq:ohrotene}) with $J''=J$ and the parameters from Table~\ref{tab:const}.  Considering electric dipole transitions only,
radiative transitions occur between adjacent $J$ levels only and we index the transition wave numbers by the lower $J$:
$\tnu_J=\Enu_{J+1}-\Enu_{J}$.  The radiative field is specified by the photon occupation numbers at the transition frequencies,
$n(\tnu_{J})$, assuming the vacuum mode density and unpolarized conditions.  If the radiation field can be characterized by a single
temperature $T_r$, the photon occupation numbers follow the Bose-Einstein statistics,
\begin{equation}
n(\tnu_{J})=[\exp(hc\tnu_J/k_BT_r)-1]^{-1} \label{eq:bose}
\end{equation}
with the Boltzmann constant $k_B$.  By the rotational transitions, the radiative field is sampled at the approximately equidistant wave
numbers $\tnu_J\approx2B_0(J+1)$.

The transition rates for radiative emission {\it out} of a level $J$ are given by
\begin{equation}
 k^{\rm em}_{J\to J-1}=A_J[1+n(\tnu_{J-1})]
\end{equation}
where $A_J$ is the Einstein coefficient for the spontaneous decay of level $J$:
\begin{equation}
  A_J=(16\pi^3/3\epsilon_0h)\tnu_{J-1}^3\mu_0^2J/(2J+1)~~(=A_{J\to J-1}).  \label{eq:einstein}
\end{equation}
from Ref.\ \cite{bernath_spectra_2005} assuming the H\"onl-London factors [Eq.\ (9.115); Table 9.4 for R transitions].  Absorption
processes {\it out} of $J$ have the rate
\begin{equation}
  k^{\rm abs}_{J\to J+1}=A_{J+1}n(\tnu_{J})(2J+3)/(2J+1).
\end{equation}
Together with the {\it in}-going processes into level $J$ the rate equation model is obtained:
\begin{equation}
  dp_J/dt =  k^{\rm abs}_{J-1\to J}p_{J-1} - (k^{\rm abs}_{J\to J+1}+k^{\rm em}_{J\to J-1})p_J + k^{\rm em}_{J+1\to J}p_{J+1}.
\label{eq:rateeq}
\end{equation}
We integrate the differential equation system with initial populations $p^0_J$ and close the system by setting $p_J\equiv0$ and
$A_{J}\equiv0$ for $J>J_{\rm max}$ as well as for $J<0$.  The used $p^0_J$ and, hence, $p_J$ are normalized ($\sum_J p_J = 1$).
Initial populations are defined by a Boltzmann distribution of temperature $T_0$.  We work at $T_0<6\,000$~K and for this upper
temperature limit, $p_J<10^{-4}$ for $J>50$.  In the model we use $J_{\rm max}=50$.

The natural lifetimes $\tau_J=1/A_J$ decrease rapidly with $J$.  As a consequence each excited rotational state cools faster than the
next lower one.  With the condition $\Delta J=\pm 1$ (R transitions, only) this slows down the radiative cascade into the rotational
ground state.  Moreover, the higher-$J$ states with shorter lifetime already thermalize among each other before population accumulates
in the lower states.  Approximately, the accumulation rate is limited by the lifetime of the lowest already thermalized level.  This
behavior significantly reduces the influence of the initial populations $p^0_J$ on those occuring after some in-vacuo relaxation time.

It should be noted that the decay constants of the $J$ levels are not in general identical to $A_J$, but modified by the ambient
radiation field.  In particular, the decay constant of the $J=1$ level, from finding the solution of Eq.~(\ref{eq:rateeq}), amounts to
$[1+4n(\tnu_{J=0})]A_1$.  Even at the present low-level cryogenic radiation field it is, thus, larger than $A_1$ by about 12\% (see
p.~\pageref{sec:radsec} below).

\vspace{10mm}
\section*{Systematic uncertainties}
The systematic uncertainties of the parameters from the fit of the probing signals are determined for various influences as follows.

{\em Starting temperature.} Fit results for the starting temperatures of 4000~K and 6000~K are compared, estimating the variations of
the fit parameters for a starting temperature difference of $\pm2000$~K around $T_0=6000$~K. This yields relative effects of $<0.1$\%
for $A_1$ and $A_2$ and of 0.15\% for $A_3$, which are negligible w.r.t.\ the $1\sigma$ statistical uncertainties (1.3\% for $A_1$ and
3\% for $A_2$ and $A_3$).  Similarly, the 0.03~K uncertainty of $T_r$ is neglected. For the fitted relative detachment cross sections,
the uncertainties due to the starting temperature, determined by the same method, are linearly added to the $1\sigma$ statistical
uncertainty to yield a total estimated uncertainty.

{\em Rotational variation of the reference photodetachment cross section.}  The effect of a $J$-dependence in $\sigma_r$ at
15\,754~\icm\ can de estimated by considering the fit to $S_4$.  At $t>150$~s, this essentially is determined by $J=1$ photodetachment
only.  For the final approach to equilibrium between $J=1$ and $J=0$, with relative populations at $t=150$~s of
$p_1\approx p_0\approx 0.5$ at $t=150$~s, the signal $S_4\approx N_{J=1}(\tnu_4,t)/[N_{J=1}(\tnu_r,t)+ N_{J=0}(\tnu_r,t)]$ can be
modeled explicitly, introducing a relative cross section difference $\eta_r=\sigma_{r}^{J=1}/\sigma_{r}^{J=0}-1$.  Fitting this model
to the data using different near-zero values of $\eta_r$ yields a sensitivity of $\Delta A_1/\Delta\eta_r=+1.45\times10^{-3}$~s$^{-1}$.
We also searched for a component describing the $J=1$ decay in the reference signal $N(\tnu_r,t)$.  This component has a decay rate of
$[1+4n(\tnu_{J=0})]A_1$ according to the radiative model.  The search yields $\eta_r=-0.01(2)$.  Hence, the systematic effect on $A_1$
due to the $J$-dependence of $\sigma_r$ is estimated as $\Delta A_1=-1.5(3.0)\times10^{-5}$~s$^{-1}$.  We choose to keep the value from
the overall fit, $A_1=5.17(7)\times10^{-3}$~s$^{-1}$, and linearly add a systematic uncertainty of $\pm1$\% due to the $J$-dependence
of $\sigma_r$.  The previous measurement in Ref.\ \cite{hlavenka_absolute_2009} indicates that the photodetachment cross section is
independent of $J$ for $J=0$ to $2$ within $\sim$10\%.  Hence, for $J=2$ and $3$, we admit a $\sigma_r$ variation of $|\eta_r|<0.1$.
Assuming the same relative effect of this variation on the Einstein coefficients, we estimate a systematic uncertainty of $\pm3$\% for
$A_2$ and $A_3$, which is of similar order of magnitude as their statistical ones.

{\em Differential laser depletion.}  Additional decays are expected in the normalized signals if photodetachment differently depletes
the $J$-levels.  An extreme upper limit for such effects is given by the observed decay constant $k_0=1.647(5)\times10^{-3}$~s$^{-1}$
of the reference signal $N(\tnu_r,t)$.  Considering the variation of $\sigma_r$ and the effect of the probing pulses at the various
wave numbers, we estimate the differential laser depletion between $J=1$ and $J=0$ to represent $<0.03k_0\sim5\times10^{-5}$~s$^{-1}$,
which introduces a relative uncertainty of $\sim$1\% for $A_1$.  A wider limit of $<0.1k_0\sim2\times10^{-4}$~s$^{-1}$ applies to
$J\geq2$, which again is of similar order of magnitude as the statistical uncertainties for these values.

{\em Overall uncertainties.}  With $\sim$1\% uncertainties for both the $J$-dependence of $\sigma_r$ and the differential laser
depletion and a $1\sigma$ relative statistical uncertainty of 1.4\%, we estimate the overall relative uncertainty of $A_1$ to 3.5\%.
For $A_2$ and $A_3$ the $1\sigma$ relative statistical uncertainties are 3.2\% and 3.1\%, respectively.  Linearly adding the mentioned
systematic uncertainty ranges, we estimate an overall relative uncertainty of 10\% for $A_2$ and 7\% for $A_3$.

{\em Electric deflection fields.}  The 60-keV OH$^-$ beam is deflected by electric fields $F$ up to 120~kV/m in the main CSR dipoles.
This mixes a rotational state $|JM\rangle$ with neighboring levels $|J\pm1\,M\rangle$ where the admixture ampitude is, from the
coupling matrix element \cite{chang_cmistark_2014},
\begin{equation}
  |\varepsilon_{J',J}^M| = \frac{\mu F}{|E_J-E_{J'}|} \frac{\sqrt{J_>^2-M^2}}{\sqrt{(2J_>+1)(2J_>-1)}}
        \leq |\varepsilon_{J',J}^0| \approx  \frac{\mu F}{2|E_J-E_{J'}|} \label{eq:smix}
\end{equation}
with $J_>=\max(J,J')$.  With $|E_J-E_{J'}|\geq 2B$ ($B=2.32\times10^{-3}$~eV) and $\mu\sim1~\rm{D}$, the interaction energy is $\mu
F\sim2.5\times10^{-6}$~eV and the admixture amplitude $|\varepsilon_{J',J}^M|\lesssim3\times10^{-4}$.  Squared matrix elements for
additional decay channels opening up by this admixture will be smaller than those of allowed decay channels by a factor of
$|\varepsilon_{J',J}^M|^2\lesssim1\times10^{-7}$, which can be neglected.  The steady-state admixed populations
$|\varepsilon_{J',J}^M|^2$ from different $J$ levels are also much smaller than the typical relevant $J$-level populations occurring in
the present experiment.

\vspace{10mm}
\section*{\boldmath \OH\ dipole moment function and vibrational averaging}
\begin{figure}[t]
\includegraphics[width=0.495\columnwidth]{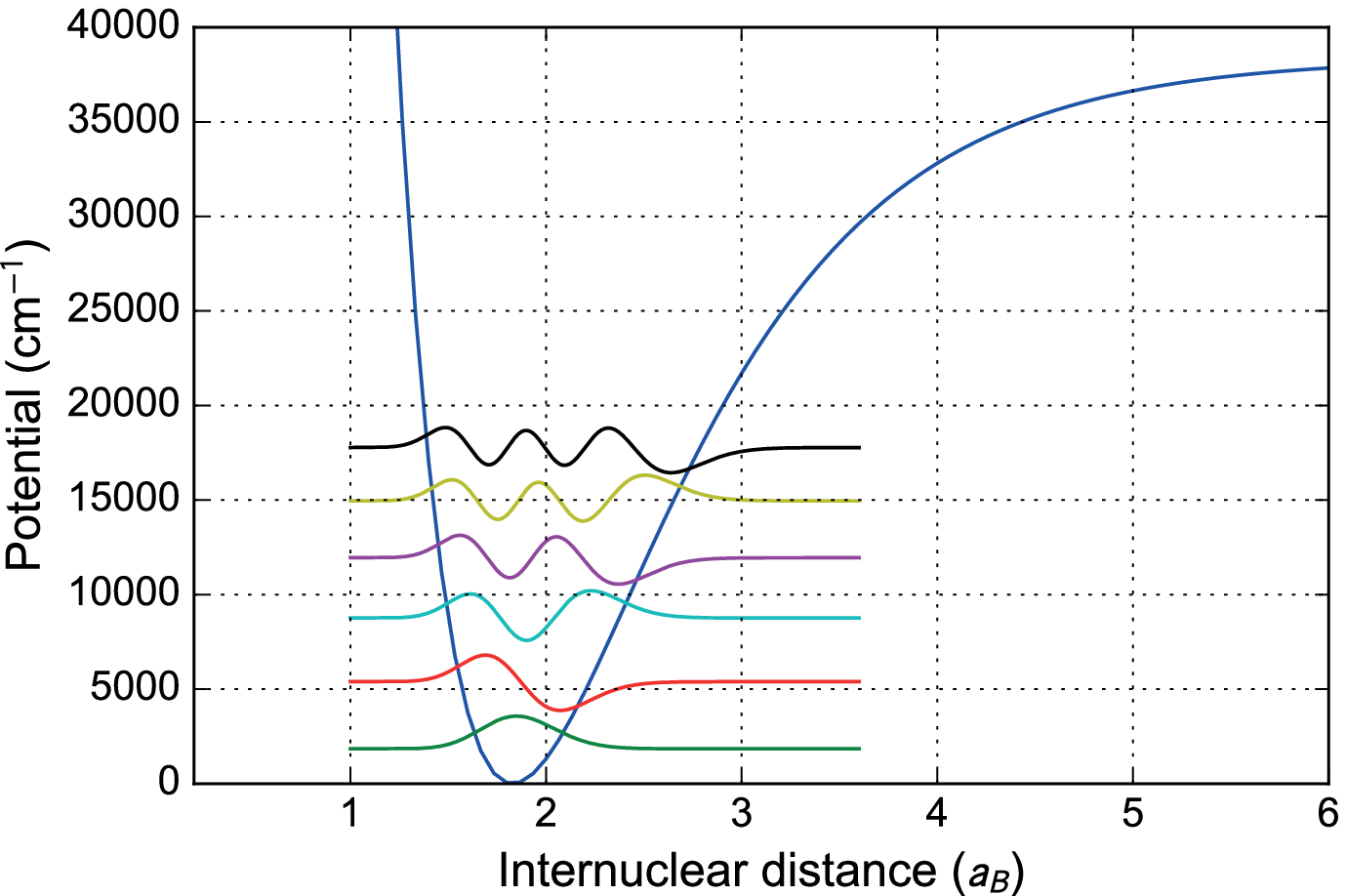}
\includegraphics[width=0.495\columnwidth]{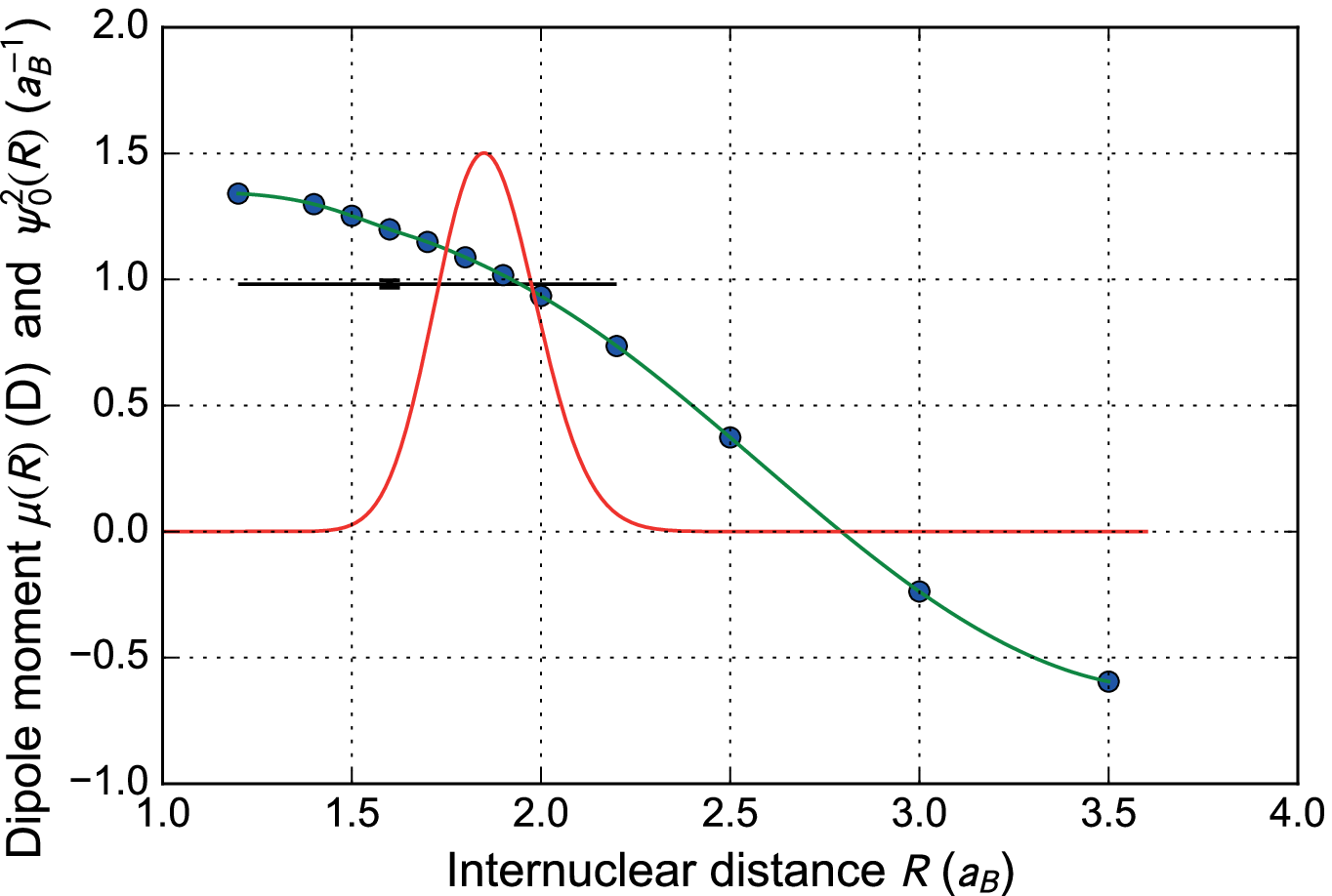}
\caption{Left: Morse potential of \OH\ (parameters according to Ref.\ \cite{bernath_spectra_2005}, p.\ 212, using $D_e=38383.3$~\icm,
  $\omega_e=3731.0$~\icm, and $r_e=1.829$~$a_B$) and $v=0$ to 5 vibrational wave functions (zero lines at the vibrational
  energies). Right: Dipole moment function $\mu(R)$ with values (dots) by \citet{werner_molecular_1983}, converted to the OH
  center-of-mass from Table~V (MCFCF-SCEP) and cubic interpolation (line).  Also shown is the $v=0$ probability distribution of the
  internuclear distance $R$ and the measured dipole moment from the main paper (horizontal line and overall
  uncertainty).\label{fig:average}}
\end{figure}

In the relaxation and probing model, the Einstein $A$ coefficients from Eq.\ (\ref{eq:einstein}) can either be fixed by giving the
molecular dipole moment $\mu_0$ of \OH\ or determined from the observed signal decays, as demonstrated in the main paper.  In the
latter case, each $A_J$ offers the option of extracting an effective molecular dipole moment $\mu_{0,J}$ (see Table~I of the main
paper).  Considering the discussion of the ro-vibrational line strength $S_{J'J''}$ in \citet{bernath_spectra_2005}, p.\ 275f., using
the H\"onl-London factors and keeping the Herman--Wallis factor $F(m)$ \cite{bernath_spectra_2005} at 1 in the expression for
$S_{J'J''}$ [Eq.\ (7.229)], the value $\mu_{0}$ discussed in the main paper is equal to the vibrationally averaged dipole moment
$M_{00}$, obtained with the vibrational wave functions for the $J=0$ molecule.  We have modeled the \OH\ vibrational wave function
using a Morse potential and calculated the average of the calculated dipole moment function $\mu(R)$ \cite{werner_molecular_1983} (see
Fig.\ \ref{fig:average}), which yields $\mu_0=M_{00}=1.037$~D.  This is lower than the value at the equilibrium internuclear distance
(Ref.\ \cite{werner_molecular_1983}, Table~IV, MCFCF(5)-SCEP) by 0.035 and still deviates by 0.055~D or $\sim$5.3\% from the measured
value, $\mu_0=(0.982 \pm 0.015)$~D.  Using the calculated $\mu_0$, the rotational lifetimes are underestimated with respect to the
measured values by $\sim$10\%..  Calculating the wave function for $J=1$ including the centrifugal potential yields an estimated
relative reduction of $\mu_{0}$ by the Herman--Wallis factor (assuming the same electronic potential for all $J$) of only
$\sim${}$2\times10^{-4}$ for the $1\to0$ transition.

\vspace{10mm}
\section*{\boldmath Radiation field at the lowest \OH\ transition}
\label{sec:radsec}
\begin{table}[b]
  \caption{Photon occupation numbers for $\tnu_{J=0}=37.47$~\icm\ and various radiation field temperatures $T_r$.\label{tab:rad}}
\begin{minipage}{10cm}
\begin{ruledtabular}
\begin{tabular}{ddl}
  \multicolumn{1}{c}{Temperature (K)} & \multicolumn{1}{c}{$n(\tnu_{J=0})$} & \multicolumn{1}{c}{Symbol}\\
  \hline
  6 & 1.252\times10^{-4} & $n_6$ \\
  15.1(1) & 2.91(9)\times10^{-2} & $\bar{n}$ \\
  300 & 5.079 &  $n_{300}$
 \end{tabular}
\end{ruledtabular}
\end{minipage}
\end{table}
In the experiment, we derive $T_r=15.1(1)$~K by sampling the radiation field on the transition $\tnu_{J=0}=37.47$~\icm.  The
corresponding photon occupation number, from Eq.\ (\ref{eq:bose}), is $n(\tnu_{J=0})=2.91(9)\times10^{-2}=\bar{n}$.  A plausible
assumption is that the effective photon occupation number $\bar{n}$ seen by the stored \OH\ ions reflects a linear superposition of
effects from the CSR vacuum chambers at a temperature near 6~K \cite{von_hahn_cryogenic_2016} and from openings towards
room-temperature surfaces (300~K).  With a fractional room-temperature influence of $\epsilon$, the effective photon occupation number
from this superposition would be $\bar{n}=\epsilon n_{300}+(1-\epsilon)n_6$.  Table~\ref{tab:rad} lists photon occupation numbers for
the radiative temperatures of 6~K (close to the measured temperature of the CSR vacuum chambers) and for a 300~K environment.  With
these values, we find $\epsilon\sim \bar{n}/n_{300}=5.7(2)\times10^{-3}$ for the fractional room-temperature influence.  While it is
difficult to estimate this value from the geometry, effective solid angles, reflection conditions, etc., this value is of the order of
magnitude of the surface area that may be affected by room-temperature openings in the present arrangement of the CSR.  In fact, a
fraction of $\epsilon$ of the CSR circumference (35~m) amounts to $\sim$20~cm, of the order of the beam tube diameter.

\vspace{10mm}
\section*{\boldmath Contamination by $^{17}$O$^-$}
In the data of Fig.~2 of the main paper, we find a background corresponding to a fraction of $(0.46\ldots1.92)\times10^{-2}$ of
the rate at the reference $\tnu_r$.  Dividing by the laser-intensity normalization factor $S_0\sim3$ (see the main paper), this yields
for the photodetachment background a fractional size of $(0.13\ldots0.45)\times10^{-2}$ compared to the reference
photodetachment rate.  We assume this reference rate to be dominated by \OH.  If we explain the background through a contamination by
$^{17}$O$^-$, the different photodetachment cross sections of \OH\ and O$^-$ must be considered.  Taking Ref.\
\cite{hlavenka_absolute_2009} and the work cited therein, the O$^-$ photodetachment cross section around 1.95~eV ($\sim$15\,751~\icm)
is smaller than that of \OH\ at the same energy by a factor of $\sim$0.75, which yields a fractional abundance of $^{17}$O$^-$ in the
stored beam of $(1.7\ldots5.9)\times10^{-3}$.  This value is plausible considering the ratio of the $^{16}$O$^-$ and
$^{16}$OH$^-$ peaks found in mass spectra of the applied sputter ion source and the $^{17}$O$^-$ natural abundance of
$3.8\times10^{-4}$ \cite{molnar_appendixes_2011}.  Variations of the value between runs can originate from changes of the ion source
parameters. Moreover, the contaminating $^{17}$O$^-$ may be partially suppressed by the dispersion of the mass selecting magnet in the
injection beamline of the CSR and the amount of the suppresion can vary with the precise tuning of the injection beam line.

\vspace{10mm}
\section*{Photodetachment cross-section ratios for rotational population probing}
For reference we give the probing wave numbers $\tnu_k$ and the experimental photodetachment cross-section ratios for \OH\ from this
work, together with the model results (Table \ref{tab:ratios}).  The measured ratios are precise enough ($\sim$3 to 16\%) to serve as
experimentally derived probing sensitivities and may be used to derive rotational population fractions from comparing photodetachment
rates at specific probing wave numbers $\tnu_k$.

\begingroup\squeezetable
\begin{table}[h]
  \caption{Cross section ratios $\sigma_J(\tnu_k)/\sigma_{J=0}(\tnu_3)$ at the probing wave numbers $\tnu_k$ using the results plotted
    in Fig.\ 3 of the main paper ($J=0$ to $3$).  Measured results are compared to the model results where
    available.  From the model, $\sigma_{J=0}(\tnu_3)/\sigma_r=0.3488$.
    \label{tab:ratios}}
\hspace*{-8mm}\begin{minipage}{1.1\columnwidth}
\begin{ruledtabular}
\begin{tabular}{rr@{}dddddddd}
  &&
  \multicolumn{2}{c}{$J=0$} & 
  \multicolumn{2}{c}{$J=1$} & 
  \multicolumn{2}{c}{$J=2$} & 
  \multicolumn{2}{c}{$J=3$} 
\\
  \multicolumn{1}{c}{$k$} & 
  \multicolumn{1}{c}{$\tnu_k$} & 
  \multicolumn{1}{c}{Exp.} &  \multicolumn{1}{c}{Model} & 
  \multicolumn{1}{c}{Exp.} &  \multicolumn{1}{c}{Model} & 
  \multicolumn{1}{c}{Exp.} &  \multicolumn{1}{c}{Model} & 
  \multicolumn{1}{c}{Exp.} &  \multicolumn{1}{c}{Model}  
\\
  \hline
1  & 14879  & 
       1.52(5) &       1.82  & 
      1.17(4)  &       1.64  & 
      1.31(12)  &      1.71  & 
 &         1.68  \\ 
2  & 14859  & 
      1.08(4) &      1.31  & 
      1.16(4)  &        1.57  & 
      1.24(10)  &       1.63  & 
 &       1.40  \\
3  & 14769  & 
      \multicolumn{2}{c}{
~~~~~~~~~~1.00\footnote{Reference value.}} & 
      0.587(21) &       0.635  & 
      0.651(29) &       0.791  & 
      0.89(12) &       0.874  \\
4  & 14732  & 
 &          0  & 
      0.603(20)  &        0.541  & 
      0.626(26)  &        0.629  & 
      0.71(9)  &       0.728  \\
5  & 14672  & 
 &          0  & 
 &          0  & 
	0.179(11) &  0.152 & 
        0.281(24)  &       0.220  \\ 
6  & 14616  & 
 &          0  & 
 &          0  & 
 &          0  & 
     0.170(27)  &        0.167  \\ 
7  & 14561  & 
 &          0  & 
 &          0  & 
 &          0  & 
      &     0.0124  \\ 
8  & 14495  & 
 &          0  & 
 &          0  & 
 &          0  & 
 &          0  \\
9  & 14428  & 
 &          0  & 
 &          0  & 
 &          0  & 
 &          0  \\
10  & 14360  & 
 &          0  & 
 &          0  & 
 &          0  & 
 &          0  \\
\end{tabular}
\end{ruledtabular}
\end{minipage}
\end{table}
\endgroup

\vspace{5mm}


\input{OH-supplement.bbl}

%% file: OH-supplement.bbl
%